\documentclass[a4paper,10pt]{article}
\usepackage{amsmath,amsthm,amsfonts,amssymb,graphicx}

\newcommand{\um}{\frac{1}{2}}
\newcommand{\oneI}{1\hspace{-0.1cm}{\rm I}}
\newcommand{\FM}{Form^{\mathcal{L}}}

\newcommand{\Prob}{{\tt Prob}}
\newcommand{\NOT}{{\tt NOT}}
\newcommand{\SNOT}{\sqrt{\tt NOT}}
\newcommand{\AND}{{\tt AND}}
\newcommand{\OR}{{\tt OR}}
\newcommand{\C}{\mathbb{C}}
\newcommand{\ket}[1]{\left\vert {#1} \right\rangle}
\newcommand{\dket}[1]{\left\Vert {#1} \right\rangle \! \rangle}

\newtheorem{theorem}{Theorem}[section]
\newtheorem{lemma}{Lemma}[section]

\newtheorem{corollary}{Corollary}[section]
\newtheorem{definition}{Definition}[section]
\newtheorem{example}{Example}[section]

\numberwithin{equation}{section}

\begin{document}

\title{\textbf{Qubit Semantics and Quantum Trees}}
\author{{\sc Maria Luisa Dalla Chiara\footnote{Dipartimento di Filosofia,
         Universit\`a di Firenze, Via Bolognese 52, 50139 Firenze, Italy.
         e-mail: {\tt dallachiara@unifi.it}}$\:$,
         Roberto Giuntini\footnote{Dipartimento di Scienze Pedagogiche
         e Filosofiche, Universit\`a di Cagliari, Via Is Mirrionis 1, 09123
         Cagliari, Italy. e-mail: {\tt giuntini@unica.it}}} \\
        {\sc Alberto Leporati$^\ddagger$, Roberto Leporini%
        \footnote{Dipartimento di Informatica, Sistemistica e Comunicazione
        (DISCo), Universit\`a degli Studi di Milano -- Bicocca, Via Bicocca
        degli Arcimboldi 8, 20126 Milano, Italy. e-mail:
        {\tt alberto.leporati@unimib.it}, {\tt leporini@disco.unimib.it}} \
        \footnote{This work has been supported by MIUR$\backslash$COFIN
        project ``Formal Languages and Automata: Theory and Applications''.}}
}
\date{}

\maketitle

\begin{abstract}
In the qubit semantics the \emph{meaning} of any sentence $\alpha$
is represented by a \emph{qu\-register}: a unit vector of the
$n$--fold tensor product $\otimes^n \C^2$, where $n$ depends on
the number of occurrences of atomic sentences in $\alpha$ (see
\cite{cdcgl-01}). The logic characterized by this semantics,
called {\it quantum computational logic\/} (QCL), is {\it
unsharp\/}, because the non-contradiction principle is violated.
We show that QCL does not admit any logical truth. In this
framework, any sentence $\alpha$ gives rise to a \emph{quantum
tree}, consisting of a sequence of unitary operators. The quantum
tree of $\alpha$ can be regarded as a quantum circuit that
transforms the quregister associated to the occurrences of atomic subformulas of
$\alpha$ into the quregister associated to $\alpha$.

\end{abstract}

\section{Introduction}

The theory of logical gates in quantum computation has suggested
the semantic characterization of a non standard form of quantum
logic, that has been called {\em quantum computational logic\/}.
We will first recall some basic notions of quantum computation.
Consider the two--dimensional Hilbert space $\C^2$ (where any
vector $\ket{\psi}$ is represented by a pair of complex numbers).
Let $\mathcal{B}^{(1)}=\{ \ket{0}, \ket{1} \}$ be the canonical
orthonormal basis for $\C^2$, where $\ket{0} = (1,0)$ and $\ket{1}
= (0,1)$.

\begin{definition}[Qubit]
    A \emph{qubit} is a unit vector $\ket{\psi}$ of the Hilbert space $\C^2$.
\end{definition}
 Recalling the Born rule, any qubit  $\ket{\psi}= c_0 \ket{0} + c_1
\ket{1}$ (with $|c_0|^2 + |c_1|^2=1$) can
 be regarded as an {\it uncertain piece
of information\/}, where the answer {\it NO\/} has probability
$|c_0|^2$, while the answer{\it YES\/} has probability $|c_1|^2$.
 The two basis-elements
$\ket{0}$ and $\ket{1}$ are usually taken as encoding the
classical bit-values $0$ and $1$, respectively. From a semantic
point of view, they can be also regarded as the classical
truth-values {\it Falsity\/} and {\it Truth\/}.

An $n$-qubit system (also called \emph{$n$--quregister} or
\emph{quantum register of size $n$}) is represented by a unit
vector in the $n$-fold tensor product Hilbert space $\otimes^n
\C^2
:= \underbrace{\C^2\otimes\ldots\otimes\C^2}_{n-times}$.
 We will use $x,y,\ldots$ as variables ranging over the set
$\{0,1\}$. At the same time, $\ket{x},\ket{y},\ldots$ will range
over the basis $\mathcal{B}^{(1)}$. Any factorized unit vector
$\ket{x_1}\otimes\ldots\otimes \ket{x_n}$ of the space $\otimes^n
\C^2$ will be called an $n$--\emph{configuration} (which can be
regarded as a quantum realization of a classical bit sequence of
length $n$). Instead of $\ket{x_1}\otimes\ldots\otimes \ket{x_n}$
we will simply write $\ket{x_1,\ldots,x_n}$. Recall that the
dimension of $\otimes^n\C^2$ is $2^n$, while the set of all
$n$--configurations
$\mathcal{B}^{(n)}=\{\ket{x_1,\ldots,x_n}:x_i\in \{0,1\}\}$ is an
orthonormal basis for the space $\otimes^n \C^2$. We will call
this set a \emph{computational basis} for the $n$--quregisters.
Since any string $x_1,\ldots,x_n$
 represents a natural number
$j\in[0,2^n-1]$ (where $j=2^{n-1} x_1+2^{n-2} x_2+\ldots+x_n$),
any unit vector of $\otimes^n \C^2$ can be shortly expressed in
the following form: $\sum_{j=0}^{2^n-1} c_j \dket{j}$, where $c_j
\in \C$, $\dket{j}$ is the $n$-configuration corresponding to the
number $j$ and $\sum_{j=0}^{2^n-1} |c_j|^2=1$.

\section{Quantum logical gates}
An $n$-input/$n$-output {\it quantum logical gate\/} is a
computation device that transforms an $n$--quregister into an
$n$--quregister. From the mathematical point of view, a quantum
logical gate can be described as a unitary operator that acts on
the vectors of the Hilbert space $\otimes^n \C^2$. We will now
introduce some examples of quantum logical gates. Since they are
described by unitary operators, it will be sufficient to determine
their behaviour on the elements of the computational basis
$\mathcal{B}^{(n)}$.

\begin{definition}[The NOT gate]
   For any $n \geq 1$, the NOT gate  is the linear operator  $\NOT^{(n)}$
   defined on $\otimes^{n} \C^2$
    such that for
   every element $\ket{x_1,\ldots,x_n}$ of the computational basis  $\mathcal{B}^{(n)}$:
   $$\NOT^{(n)}(\ket{x_1,\ldots,x_n}) = \ket{x_1,\ldots,x_{n-1}} \otimes
   \ket{1-x_n}.$$
\end{definition}

In other words, $\NOT^{(n)}$ inverts the value of the last element
of any basis--vector of $\otimes^{n} \C^2$.

\begin{definition}[The Petri-Toffoli  gate]
   For any $n \geq 1$ and any $m \geq 1$ the Petri-Toffoli gate is the
linear operator $T^{(n,m,1)}$ defined on $\otimes^{n+m+1} \C^2$
such that
   for every element $\ket{x_1,\ldots,x_n}\otimes\ket{y_1,\ldots,y_m}
   \otimes\ket{z}$ of the computational basis $\mathcal{B}^{(n+m+1)}$:
   $$
   T^{(n,m,1)}(\ket{x_1,\ldots,x_n}\otimes\ket{y_1,\ldots,y_m}
   \otimes\ket{z}) = \ket{x_1,\ldots,x_n} \otimes \ket{y_1,\ldots,y_m} \otimes
   \ket{x_n y_m \oplus z},$$
    where $\oplus$ represents the sum modulo $2$.
\end{definition}

One can easily show that both $\NOT^{(n)}$ and $T^{(n,m,1)}$ are
unitary operators.

The gate $T^{(n,m,1)}$ is very similar to a gate introduced by
Petri in \cite{Pe-67}. For $n=m=1$ we obtain the well known
Toffoli gate (\cite{To-80}), which is essentially identical to
Feynman's {\em Controlled-Controlled-NOT gate}. Both classical
conjunction and classical negation are realized by this gate in a
reversible way.

The quantum logical gates we have considered so far are, in a
sense, ``semiclassical". A quantum logical behaviour only emerges
in the case where our gates are applied to superpositions. When
restricted to classical registers, such operators turn out to
behave as classical truth-functions. We will now consider a {\it
genuine quantum gate\/} that transforms classical registers
(elements of $\mathcal{B}^{(n)}$) into quregisters that are
superpositions.

\begin{definition}[The square-root-of-NOT gate]
   For any $n \geq 1$, the square-root-of-NOT  is the linear operator  $\sqrt{\NOT}^{(n)}$
   defined on $\otimes^{n} \C^2$
    such that for
   every element $\ket{x_1,\ldots,x_n}$ of the computational basis  $\mathcal{B}^{(n)}$:
   $$\sqrt{\NOT}^{(n)}(\ket{x_1,\ldots,x_n}) = \ket{x_1,\ldots,x_{n-1}}
   \otimes \um((1+i) \ket{x_n} + (1-i) \ket{1-x_n}).$$
\end{definition}
One can easily show that $\sqrt{\NOT}^{(n)}$ is a unitary
operator. The basic property of $\sqrt{\NOT}^{(n)}$ is the
following: $$ \text{for any}\,\,\, \ket{\psi} \in \otimes^{n}
\C^2, \,\,\,\sqrt{\NOT}^{(n)}(\sqrt{\NOT}^{(n)}(\ket{\psi}))= \NOT^{(n)}
(\ket{\psi}). $$ In other words, applying twice the square root of
the negation means negating.

Interestingly enough, the square-root-of-NOT gate has some
physical models and implementations. As an example, consider an
idealized atom with a single electron and two energy levels: a
\emph{ground state} (identified with $\ket{0}$) and an
\emph{excited state} (identified with $\ket{1}$). By shining a
pulse of light of appropriate intensity, duration and wavelength,
it is possible to force the electron to change energy level. As a
consequence, the state (bit) $\ket{0}$ is transformed into the
state (bit) $\ket{1}$, and viceversa: $\ket{0} \mapsto \ket{1};
\quad \ket{1} \mapsto \ket{0}$. We have thus obtained a typical
physical model for the gate $\NOT^{(1)}$.

Now, by using a light pulse of half the duration as the one needed to perform
the NOT operation, we effect a half--flip between the two logical states.
The state of the atom after the half pulse is neither $\ket{0}$ nor $\ket{1}$,
but rather a superposition of both states: $\ket{0} \mapsto \frac{1+i}{2}
\ket{0} + \frac{1-i}{2} \ket{1}; \quad \ket{1} \mapsto \frac{1-i}{2} \ket{0}
+ \frac{1+i}{2} \ket{1}$.

As expected, the square-root-of NOT gate has no Boolean
counterpart.
\begin{lemma}
   There is no function $f: \{0,1\} \to \{0,1\}$
   such that for any  $x\in~\{0,1\}: \, f(f(x))= 1-x$.
\end{lemma}
\begin{proof}
Suppose, by contradiction, that such a function $f$ exists.
Two cases are possible:
(i)\space $f(0)=0$; (ii)\space $f(0)=1$.
\par\noindent
 (i)\space By hypothesis,
$f(0)=0$. Thus, $1=f(f(0))=f(0)=0$, contradiction.
\par\noindent
(ii)\space By hypothesis, $f(0)=1$. Thus, $1=f(f(0))=f(1)$. Hence,
$f(0)=f(1)$. Therefore, $1=f(f(0))=f(f(1))=0$, contradiction.
\end{proof}

Interestingly enough,$\SNOT$ does not have even any fuzzy
counterpart.
\begin{lemma}
   There is no continuous function $f:[0,1] \to [0,1]$ such that for any  $x
   \in~[0,1]: \,f(f(x))=1-x$.
\end{lemma}
\begin{proof}
Suppose, by contradiction, that such a function $f$ exists. First,
we prove that $f(\frac{1}{2})=\frac{1}{2}$. By hypothesis,
$f(f(\frac{1}{2}))=1-\frac{1}{2}=\frac{1}{2}$. Hence,
$f(f(f(\frac{1}{2})))=f(\frac{1}{2})$. Thus,
$1-f(\frac{1}{2})=f(\frac{1}{2})$. Therefore,
$f(\frac{1}{2})=\frac{1}{2}$. Consider now $f(0)$. One can easily
show: $f(0)\not=0$ and $f(0)\not=1$. Clearly,
$f(0)\not=\frac{1}{2}$ since otherwise we would obtain
$1=f(f(0))=f(\frac{1}{2})=\frac{1}{2}$. Thus, only two cases are
possible: (i)\space $0<f(0)<\frac{1}{2}$; (ii)\space
$\frac{1}{2}<f(0)<1$.
\par\noindent
(i)\space By hypothesis, $0<f(0)<\frac{1}{2}<1=f(f(0))$. Consequently, by
continuity, $\exists x\in(0,f(0))$ such that $\frac{1}{2}=f(x)$.
Accordingly, $\frac{1}{2}=f(\frac{1}{2})=f(f(x))=1-x$. Hence,
$x=\frac{1}{2}$, which contradicts $x<f(0)<\frac{1}{2}$.
\par\noindent
(ii)\space By hypothesis,
$f(\frac{1}{2})=\frac{1}{2}<f(0)<1=f(f(0))$. By continuity, $\exists
x\in(\frac{1}{2},f(0))$ such that $f(x)=f(0)$. Thus,
$1-x=f(f(x))=f(f(0))=1$. Hence, $x=0$, which contradicts
$x>\frac{1}{2}$.
\end{proof}

Consider now the set $\bigcup_{n=1}^{\infty} \otimes^n \C^2$
(which contains all quregisters $\ket{\psi}$ ``living" in
$\otimes^n \C^2$, for a given $n\ge1$). The gates $\NOT$, $\SNOT$
and $T$ can be uniformly defined on this set in the expected way:
\begin{align*}
&\NOT(\ket{\psi}) := \NOT^{(n)}(\ket{\psi}), \hspace{3.5cm} \text{if} \;
 \ket{\psi} \in \otimes^n \C^2 \\
&\SNOT(\ket{\psi}) := \SNOT^{(n)}(\ket{\psi}), \hspace{2.9cm} \text{if} \;
 \ket{\psi} \in \otimes^n \C^2 \\
&T(\ket{\psi} \otimes \ket{\varphi} \otimes \ket{\chi}) :=
 T^{(n,m,1)}(\ket{\psi}\otimes \ket{\varphi} \otimes \ket{\chi}), \\
&\hspace{3cm}\text{if} \; \ket{\psi} \in \otimes^n \C^2, \; \ket{\varphi} \in \otimes^m \C^2
\; \text{and} \; \ket{\chi} \in \C^2
\end{align*}

On this basis, a conjunction $\AND$ and a disjunction $\OR$ can be
defined for any pair of quregisters $\ket{\psi}$ and
$\ket{\varphi}$:
$$
\AND(\ket{\psi},\ket{\varphi}):=
T(\ket{\psi}\otimes\ket{\varphi}\otimes\ket{0}).
$$
$$
\OR(\ket{\psi},\ket{\varphi}):=
\NOT(\AND(\NOT(\ket{\psi}),\NOT(\ket{\varphi}))).
$$

Clearly, $\ket{0}$ represents an ``ancilla" in the definition of
$\AND$. We will use $\AND$ ($\OR$) as a metalinguistic
abbreviation for the corresponding {\em definiens\/}.

One can easily verify that, when applied to classical bits,
$\NOT$, $\AND$ and $\OR$ behave as the standard Boolean
truth-functions.

We will now introduce the concept of {\it probability--value of a
quregister}, which will play an important role in the quantum
computational semantics. For any integer $n\ge 1$, let us first
define a particular set of natural numbers:

$$ C_1^{(n)}:=\{i \,:\, \dket{i} = \ket{x_1,\ldots,x_n} \text{and
} x_n = 1 \}. $$

Apparently, $C_1^{(n)}$ contains precisely all the odd numbers in
$[0,2^n-1]$.

\begin{definition}[Probability--value]
    Let $\ket{\psi}=\sum_{j=0}^{2^n-1} c_j \dket{j}$ be any quregister
     of
   $\otimes^n\C^2$.
   The {\em probability-value of} $\ket{\psi}$ is the real value
   $\Prob(\ket{\psi}) := \sum_{j \in C_1^{(n)}} |c_j|^2$.
\end{definition}

From an intuitive point of view, $\Prob(\ket{\psi})$ represents
the probability that the quregister $\ket{\psi}$ (which is a
superposition) {\it collapses\/} into an $n$-configuration whose
last element is $1$.
\begin{theorem}\label{th:radice}
 Let $\ket{\psi}$  and
  $\ket{\varphi}$ be two quregisters. The following
properties hold:
\begin{enumerate}\item[]
   \begin{enumerate}
  \item[\rm(i)]\quad $\Prob(\AND(\ket{\psi},\ket{\varphi}))=\Prob(\ket{\psi})\Prob(\ket{\varphi})$;
  \item[\rm(ii)]\quad $\Prob(\NOT(\ket{\psi}))=1-\Prob(\ket{\psi})$;
  \item[\rm(iii)]\quad $\Prob(\OR(\ket{\psi},\ket{\varphi}))=\Prob(\ket{\psi})+\Prob(\ket{\varphi})
                        -\Prob(\ket{\psi})\Prob(\ket{\varphi})$;
    \item[\rm(iv)]\quad Let  $\ket{\psi}=\sum_{j=0}^{2^n-1} a_j \dket{j}$.
    Then \item[]\quad
      $\Prob(\sqrt{\NOT}(\ket{\psi}))=\sum_{j\in C_1^{(n)}}\left| \dfrac{1-i}{2}c_{j-1}
    +\dfrac{1+i}{2}c_j\right |^2$;
 \item[\rm(v)]\quad
 $\Prob(\sqrt{\NOT}(\,\NOT(\ket{\psi})))=\Prob(\NOT(\sqrt{\NOT}(\ket{\psi})))$;
 \item[\rm(vi)]\quad
 Let $\ket{\psi}=\sum_{j=0}^{2^n-1} a_j \dket{j}\ket{x_j}$. Then
 $\Prob(\sqrt{\NOT}(\ket{\psi}))=\um$;
      \item[\rm(vii)]\quad $\Prob(\sqrt{\NOT}(\AND(\ket{\psi},\ket{\varphi})))=\frac{1}{2}$.
\end{enumerate}
  \end{enumerate}
\end{theorem}
\begin{proof}
\begin{align*}
&\textrm{(i)--(v)} \: \cite{DGL-03}; \\[-0.4cm] &\textrm{(vi) }
\Prob(\sqrt{\NOT}(\ket{\psi}))
=\Prob\left(\sqrt{\NOT}\left(\sum_{j=0}^{2^n-1} a_j
\dket{j}\ket{x_j}\right)\right)\\
  &=\Prob\left(\sum_{j=0}^{2^n-1} a_j
        \dket{j}\otimes\left(\um(1+i)\ket{x_j}+\um(1-i)\ket{1-x_j}\right)\right) \\
&=\Prob\left(\sum_{j=0}^{2^n-1} a_j \um
\left(1-i(-1)^{x_j}\right)\dket{j}\ket{1} +\sum_{j=0}^{2^n-1} a_j
\um \left(1+i(-1)^{x_j}\right)\dket{j}\ket{0} \right)\\
&=\sum_{j=0}^{2^n-1} \left| a_j \um \left(1-i(-1)^{x_j}\right)
\right|^2 =\sum_{j=0}^{2^n-1} |a_j|^2 \left|\um
\left(1-i(-1)^{x_j}\right) \right|^2 =\um \sum_{j=0}^{2^n-1}
|a_j|^2=\um;\\ &\textrm{(vii)}\text{
}\AND(\ket{\psi},\ket{\varphi})\text{ has the form }
\sum_{j=0}^{2^{n+m}-1} a_j \dket{j}\ket{x_j}.\\ &\text{ Thus, by
}\textrm{(vi), }
\Prob(\sqrt{\NOT}(\AND(\ket{\psi},\ket{\varphi})))=\um.\\[-1.3cm]
\end{align*}
\end{proof}

\section{Quantum computational semantics}

The starting point of the quantum computational semantics is quite
different from the standard quantum logical approach. The basic
idea is that every sentence $\alpha$ is semantically interpreted
as a quregister. From an intuitive point of view, one can say that
the meaning of a sentence is identified with the \emph{information
quantity} encoded by the sentence under consideration.

Consider a sentential language $\mathcal{L}$ with the following connectives:
\emph{negation} ($\lnot$), \emph{square root of not} ($\sqrt{\lnot}$),
\emph{conjunction} ($\land$).
Let $\FM$ be the class of all sentences of the language $\mathcal{L}$.
We will use the following metavariables: $p,q,\ldots$ for atomic sentences and
$\alpha,\beta,\ldots$ for sentences.

The basic concept of our semantics is represented by the notion of
\emph{quantum computational model}: an interpretation of the
language $\mathcal{L}$ that associates a quregister to any
sentence $\alpha$.

\begin{definition}[Quantum computational model]
   A \emph{quantum computational model} of $\mathcal{L}$ is a function
   $\emph{Qub}: Form^\mathcal{L} \to \bigcup_{n=1}^{\infty} \otimes^n \C^2$
   that associates to any sentence $\alpha$ of the language a quregister: \\
   $Qub(\alpha) :=
        \begin{cases}
          \text{a qubit}
                    & \text{if}\,\, \alpha\,\, \text{ is an atomic sentence}; \\
          \NOT(Qub(\beta))

                    & \text{if} \,\, \alpha = \lnot
                    \beta; \\

                     \SNOT(Qub(\beta))
                    & \text{if} \,\, \alpha = \sqrt{\lnot}
                    \beta; \\
                     \AND(Qub(\beta),Qub(\gamma))
                    & \text{if} \,\, \alpha = \beta \land \gamma.
        \end{cases}$
\end{definition}

We will call $Qub(\alpha)$ the \emph{information-value} of
$\alpha$. Instead of $Qub(\alpha)$, we will also write
$\ket{\alpha}_{Qub}$ (or simply $\ket{\alpha}$). Our definition
univocally determines, for any $Qub$ and any sentence $\alpha$,
the Hilbert space $\otimes^n \C^2$ to which $\ket{\alpha}_{Qub}$
belongs. Apparently, $n$ is the number of all occurrences of
atomic sentences and of the connective $\land$ in $\alpha$.
According to the intended physical interpretation, $Qub$ will
associate to each occurrence of one and the same atomic subformula
$p$ of $\alpha$ the state $\ket{p}$, that corresponds to an {\em
identical preparation\/} of the quantum system.

We can now define the notion of \emph{truth}, \emph{logical truth},
\emph{consequence} and \emph{logical consequence}.

\begin{definition}[Truth and logical truth]
   A sentence $\alpha$ is \emph{true in a quantum computational model}
   $Qub$
   (abbreviated as $\models_{Qub} \alpha$) iff
   $\Prob(Qub(\alpha))=1$;
   $\alpha$ is a \emph{logical truth} ($\models \alpha$) iff for any $Qub$,
   $\models_{Qub} \alpha$.
\end{definition}

\begin{definition}[Consequence in $Qub$ and logical consequence]
   A sentence $\beta$ is a \emph{consequence in a quantum computational
    model} $Qub$ of a  sentence $\alpha$ ($\alpha \models_{Qub} \beta$)
   iff  $\,\Prob(Qub(\alpha)) \leq
   \Prob(Qub(\beta))$;
   $\beta$ is a \emph{logical consequence} of $\alpha$ ($\alpha \models \beta$) iff
  for any $Qub$, $\,\alpha\models_{Qub}\beta$.
\end{definition}

The logic characterized by this semantics has been termed {\it
quantum computational logic} (QCL, for short)\cite{cdcgl-01}. The
following theorem shows that this logic is
  completely different from the well known {\em orthomodular quantum logic\/}
  (OQL), which is semantically characterized by the class of all
  orthomodular lattices.

\begin{theorem}
   QCL and OQL are not comparable.
\end{theorem}
\begin{proof}
        (i)\space OQL is not a sublogic of QCL. This follows from
        the fact that the idempotence property ($\alpha \models \alpha \land \alpha$) holds
in OQL, whereas it is violated in QCL. Take for example, $\ket{\alpha} =
\frac{1}{\sqrt{2}} (\ket{0}
+ \ket{1})$. Then, $\Prob(\ket{\alpha})=\frac{1}{2} > \frac{1}{4} =
\Prob(\ket{\alpha \land \alpha})$.
\par\noindent
(ii)\space QCL is not a sublogic of OQL. This follows from the fact that
the strong distributivity property
($\alpha \land (\beta \lor
\gamma) \models (\alpha \land \beta) \lor (\alpha \land \gamma)$) is violated in OQL
 (\cite{DG02}), whereas it holds in QCL. In fact, by Theorem
 \ref{th:radice}~(i)-iii)), we obtain
\begin{align*}
\Prob(\ket{\alpha\land(\beta\lor\gamma)}) &=
     \Prob(\AND(\ket{\alpha},\OR(\ket{\beta},\ket{\gamma}))) \\
        &=\Prob(\ket{\alpha})\Prob(\ket{\beta}) +
              \Prob(\ket{\alpha})\Prob(\ket{\gamma}) \\
                &\hspace{3.4cm} - \Prob(\ket{\alpha})\Prob(\ket{\beta})
                     \Prob(\ket{\gamma}) \\
                        &\le \Prob(\ket{\alpha})\Prob(\ket{\beta}) +
                         \Prob(\ket{\alpha})\Prob(\ket{\gamma}) \\
                            &\hspace{3.4cm} - \Prob(\ket{\alpha})^2\Prob(\ket{\beta})
                                  \Prob(\ket{\gamma}) \\
                                   &=\Prob(\OR(\AND(\ket{\alpha},\ket{\beta}),\AND(\ket{\alpha},
                                    \ket{\gamma}))) \\
                                       &=\Prob(\ket{(\alpha\land\beta)\lor(\alpha\land\gamma)})\\[-1.4cm]
                                       \end{align*}
\end{proof}

The logic QCL turns out to be {\it unsharp}, because the
non--contradiction principle can be violated: the negation of a
contradiction ($\lnot(\alpha \land \lnot \alpha)$) is not
necessarily true \cite{cdcgl-01}.

\begin{theorem}\label{th:threevalue}
   Let Qub be any quantum computational model and let $\alpha$ be any sentence.
   If $\Prob(Qub(\alpha)) \in \{ 0,1 \}$, then there is an atomic subformula
    $p$ of $\alpha$
   such that
   $\Prob(Qub(p)) \in \{ 0,\um,1 \}$.
\end{theorem}
\begin{proof}
Suppose that $\Prob(Qub(\alpha)) \in \{ 0,1 \}$. The proof is by
induction on the logical complexity of $\alpha$.
\par\noindent
(i)\space $\alpha$ is an atomic sentence. The proof is
trivial.
\par\noindent
(ii)\space $\alpha = \lnot \beta$.
By Theorem \ref{th:radice}(ii),  $\Prob(Qub(\alpha))=1-\Prob(Qub(\beta)) \in \{ 0,1 \}$.
The conclusion follows
by induction hypothesis.
\par\noindent
(iii)\space $\alpha = \sqrt{\lnot} \beta$.
By hypothesis and by Theorem \ref{th:radice}(vii), $\beta$ cannot be a conjunction.
Consequently, only the
following cases are possible:
(iiia)\space $\beta = p$; (iiib)\space $\beta = \lnot \gamma$; (iiic)\space $\beta =
\sqrt{\lnot} \gamma$.
\par\noindent
(iiia)\space  $\beta = p$. By hypothesis, $\Prob(\sqrt{\lnot}\beta)\in\{0,1\}$. Hence,
$\SNOT (Qub(p))=c\ket{x}$, where $\ket{x}\in\{\ket{0},\ket{1} \}$
and $|c|=1$.
We have:  $\NOT (Qub(p))= \SNOT( \SNOT (Qub(p)))=\SNOT (c\ket{x})$. By
Theorem \ref{th:radice}(iv), $\Prob(\SNOT (c\ket{x}))=\frac{1}{2}$.
As a consequence, $\Prob(Qub(\neg p))=\frac{1}{2}=\Prob(Qub(p))$.
\par\noindent
(iiib)\space $\beta = \lnot \gamma$. By Theorem \ref{th:radice}(v),
$\Prob(Qub(\sqrt{\lnot} \lnot \gamma))= \Prob(Qub(\lnot
\sqrt{\lnot} \gamma))=1-\Prob(Qub(\sqrt{\lnot} \gamma))$.
The conclusion follows by induction
hypothesis.
\par\noindent
(iiic)\space $\beta = \sqrt{\lnot} \gamma$. Then $\Prob(Qub(\sqrt{\lnot}
\sqrt{\lnot} \gamma))= \Prob(Qub(\lnot
\gamma))=1-\Prob(Qub(\gamma))$. The conclusion follows by
induction hypothesis.
\par\noindent
(iv)\space $\alpha = \beta \land \gamma$. By Theorem
\ref{th:radice}(i),
$\Prob(Qub(\beta\land\gamma))=\Prob(Qub(\beta)) \Prob(Qub(\gamma)) \in \{ 0,1 \}$. The
conclusion follows by induction hypothesis.
\end{proof}

A remarkable property of QCL is asserted by the following
Corollary of Theorem \ref{th:threevalue}.
\begin{corollary}
   There exists no quantum computational logical truth.
\end{corollary}
\begin{proof}
Suppose, by contradiction, that $\alpha$ is a logical truth. Let
$p_1,\ldots ,p_n$ be the atomic sentences occurring in $\alpha$
and let $Qub$ be a quantum computational model such that for any $i$
($1 \le i \le n$), $\Prob(Qub(p_i))\notin
\{0,\frac{1}{2},1\}$. Then, by Theorem~\ref{th:threevalue}, $\Prob(Qub(\alpha)) \notin
\{0,1\}$, contradiction.
\end{proof}

\section{Quantum trees}

For the sake of technical simplicity we slightly modify our
language. The new language contains a privileged atomic sentence
$f$ (representing {\it the falsity}) and three primitive
connectives: the negation $\lnot$, the square root of the negation
$\sqrt{\lnot}$ and a ternary conjunction $\bigwedge$.
 The connective $\bigwedge$ represents  a conjunction whose form is
``close'' to the Petri-Toffoli gate. For any sentences $\alpha$
and $\beta$ the expression $\bigwedge(\alpha,\beta,f)$ is a
sentence of the language. The usual conjunction $\alpha \land
\beta$ is dealt with as meta\-linguistic abbreviation for the
ternary conjunction $\bigwedge(\alpha, \beta,f)$. Semantically, we
will require that for any $Qub$: $$ Qub(f)=\ket{0};\,\,\,
Qub(\bigwedge(\alpha,\beta,f))=
T(Qub(\alpha)\otimes Qub(\beta)\otimes Qub(f)).$$

\begin{definition}[The Atomic Complexity of $\alpha$]
   \hspace{-0.15cm}The atomic complexity of a sentence $\alpha$ \hspace{-0.1cm}$(Atcompl(\alpha))$
   \hspace{-0.1cm}is the number of occurrences of atomic sentences in $\alpha$.
\end{definition}
For example, if $\alpha=p \land \lnot p=\bigwedge(p,\lnot p,f)$, then
$Atcompl(\alpha)=3$.

\begin{lemma}
   Let $Atcompl(\alpha)=n$.
   Then $\forall Qub: \, Qub(\alpha) \in \otimes^n \C^2$.
\end{lemma}

Hence, the space of all possible qubit--meanings of $\alpha$ is
determined by the atomic complexity of $\alpha$.

We will first introduce the notion of {\it syntactical tree} of a
sentence $\alpha$ (abbreviated as $STree^{\alpha}$). Consider all subformulas of
$\alpha$.


Any subformula may be:
\begin{itemize}
   \item an {\it atomic\/} sentence $p$ (possibly $f$);
   \item a {\it negated\/}  sentence $\lnot \beta$;
   \item a {\it square-root-negated\/} sentence $\sqrt{\lnot}
   \beta$;
   \item a {\it conjunction\/}  $\bigwedge(\beta,\gamma,f)$.
\end{itemize}

The intuitive idea of {\it syntactical tree\/} can be illustrated
as follows. Every occurrence of a subformula of $\alpha$ gives
rise to a {\it node} of $STree^{\alpha}$. The tree consists of a
finite number of {\it levels} and each level is represented by a
sequence of subformulas of $\alpha$:
\vspace{-0.3cm}
\begin{gather*}
Level_k(\alpha)\\[-0.2cm]
\vdots\\[-0.1cm]
Level_1(\alpha).
\end{gather*}
The {\it root-level} (denoted by
$Level_1(\alpha)$) consists of $\alpha$. From each node of the
tree at most 3 edges may branch according to the following {\it
branching-rule}:

\begin{figure}[ht]
\begin{center}
\includegraphics[width=7.5cm]{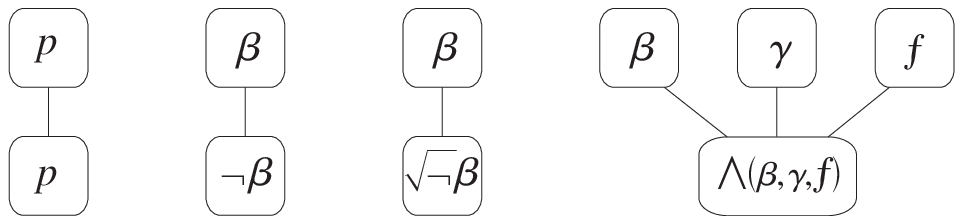}
\end{center}
\label{fig:branching-rules}
\end{figure}
\vspace{-0.5cm}
The second level ($Level_2(\alpha)$) is the sequence of
subformulas of $\alpha$ that is obtained by applying the
branching-rule to $\alpha$. The third level ($Level_3(\alpha)$) is
obtained by applying the branching-rule to each element (node) of
$Level_2(\alpha)$, and so on. Finally, one obtains a level
represented by the sequence of all atomic occurrences of $\alpha$.
This represents the {\it last level\/} of $STree^{\alpha}$. The
{\it height\/} of $Stree^{\alpha}$ (denoted by $Height(\alpha)$)
is then defined as the number of levels of $STree^{\alpha}$.

 A more formal definition of {\it
syntactical tree\/} can be given by using some standard
graph-theoretical notions.

\begin{example}
   The syntactical tree of
   $\alpha = \lnot p \land (q \land \sqrt{\lnot} p)$  is the following:

  \begin{center}
     \includegraphics[width=8.5cm]{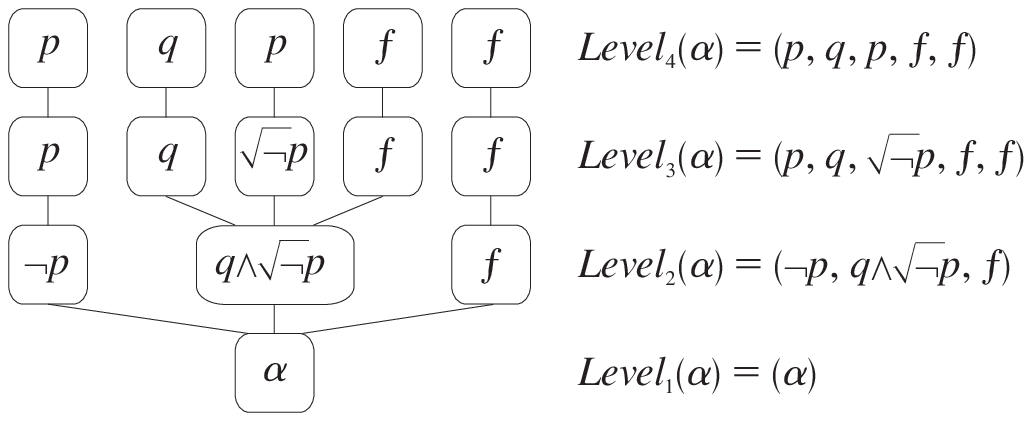}
  \end{center}
   Clearly the height of $Stree^{\alpha}$ is 4.
\end{example}
\vspace{-0.2cm}

For any choice of a quantum computational model $Qub$, the
syntactical tree of $\alpha$ determines a corresponding sequence
of quregisters. Consider a sentence $\alpha$ with $n$ atomic
occurrences ($p_1,\ldots,p_n$). Then $Qub(\alpha) \in \otimes^n
\C^2$. We can associate a quregister $\ket{\psi_i}$ to each
$Level_i(\alpha)$ of $Stree^{\alpha}$ in the following way.
Suppose that: $$Level_i(\alpha)=(\beta_1, \ldots, \beta_r).$$ Then:
$$\ket{\psi_i}=Qub(\beta_1) \otimes \ldots \otimes
Qub(\beta_r)$$ Hence: $$\begin{cases}
    \ket{\psi_1}=Qub(\alpha) \\
    \vdots \\
    \ket{\psi_{Height(\alpha)}}=Qub(p_1) \otimes \ldots \otimes Qub(p_n)
\end{cases}$$
where all $\ket{\psi_i}$ belong to the same space $\otimes^n \C^2$.

From an intuitive point of view, $\ket{\psi_{Height(\alpha)}}$ can
be regarded as a kind of {\it epistemic state\/}, corresponding to
the input of a computation, while $\ket{\psi_1}$ represents the
output.

We obtain the following correspondence:
\begin{align*}
   \mbox{$Level_{Height(\alpha)}(\alpha)$} & \mbox{$\leftrightsquigarrow
         \ket{\psi_{Height(\alpha)}}$: the input} \\
   \mbox{$\ldots$} & \mbox{$\leftrightsquigarrow \ldots$} \\
   \mbox{$Level_1(\alpha)$} & \mbox{$\leftrightsquigarrow
        \ket{\psi_1}$: the output}
\end{align*}

The notion of {\it quantum tree\/} of a sentence $\alpha$
 ($QTree^{\alpha})$ can be now defined as a particular sequence of
unitary operators that is uniquely determined by the syntactical
tree of $\alpha$. As we already know, each $Level_i(\alpha)$  of
$STree^{\alpha}$ is a sequence of subformulas of $\alpha$. Let
$Level_i^j(\alpha)$ represent the $j$-th element of
$Level_i(\alpha)$. Each node $Level_i^j(\alpha)$ (where $1 \le i
< Height(\alpha)$) can be naturally associated to a unitary
operator $Op^j_i$, according to the following {\it
operator-rule\/}:
\[
Op^j_i :=
 \begin{cases}
  \oneI^{(1)} & \text{if $Level_i^j(\alpha)$ is an atomic sentence}; \\
    \NOT^{(r)} & \text{if $Level_i^j(\alpha) = \lnot \beta$ and
    $\ket{\beta} \in \otimes^r \C^2$}; \\
      \SNOT^{(r)} & \text{if $Level_i^j(\alpha) = \sqrt{\lnot} \beta$ and
      $\ket{\beta} \in \otimes^r \C^2$}; \\
        T^{(r,s,1)} & \text{if $Level_i^j(\alpha) =
           \bigwedge(\beta,\gamma,f)$, $\ket{\beta} \in \otimes^r \C^2$ and
            $\ket{\gamma} \in \otimes^s \C^2$},
                 \end{cases}
            \]
         where $ \oneI^{(1)}$ is the identity operator on $\C^2$.

On this basis, one can associate an operator $U_i$ to each
    $Level_i(\alpha)$ (such that $1 \le
i < Height(\alpha)$):
$$
U_i := \bigotimes_{j=1}^{|Level_i(\alpha)|}
Op_i^j,
$$
    where $|Level_i(\alpha)|$ is the length of the sequence
    $Level_i(\alpha)$.

    Being the tensor product of unitary operators, every $U_i$
    turns out to be a unitary operator. One can easily show that
    all $U_i$ are defined in the same space $\otimes^n \C^2$,
    where $n$ is the atomic complexity of $\alpha$.

The notion of {\it quantum tree\/} of a sentence can be now
defined as follows.
\begin{definition}[The quantum tree of $\alpha$] The {\it quantum
tree\/} of $\alpha$ (denoted by $QTree^{\alpha}$) is the
operator-sequence $(U_1,\ldots,U_{Height(\alpha)-1})$ that is
uniquely determined by the syntactical tree of $\alpha$.
\end{definition}

As an example, consider the following sentence: $\alpha=p \land
\lnot p = \bigwedge(p,\lnot p,f)$. The syntactical tree of $\alpha$
is the following:
\vspace{-0.3cm}
\begin{align*}
\mbox {$Level_1(\alpha)=\bigwedge(p,\lnot p,f)$}\\
\mbox {$Level_2(\alpha)= (p,\lnot p,f)$}\\
\mbox {$Level_3(\alpha)= (p,p,f).$}
\end{align*}
In order to construct the quantum tree of $\alpha$, let us first
determine the operators $Op^j_i$ corresponding to each node of
$Stree^{\alpha}$. We will obtain:
\vspace{-0.2cm}
\begin{itemize}
\item $Op^1_1=T^{(1,1,1)}$, because $\bigwedge(p,\lnot
p,f)$ is connected with $(p,\lnot p,f)$ (at $Level_2(\alpha)$);
   \item $Op_2^1=\oneI^{(1)}$, because $p$ is connected with $p$ (at
         $Level_3(\alpha)$);
   \item $Op_2^2=\NOT^{(1)}$, because $\lnot p$ is connected with $p$ (at
         $Level_3(\alpha)$);
   \item $Op_2^3=\oneI^{(1)}$, because $f$ is connected with $f$ (at
         $Level_3(\alpha)$).
\end{itemize}

The quantum tree of $\alpha$ is represented by the
operator-sequence $(U_1,U_2)$, where:
\vspace{-0.55cm}
\begin{align*}
&U_1= Op_1^1=T^{(1,1,1)};\\
&U_2= Op_2^1 \otimes Op_2^2 \otimes
Op_2^3=\oneI^{(1)} \otimes \NOT^{(1)} \otimes \oneI^{(1)}.
\end{align*}

Apparently, $QTree^{\alpha}$ is independent of the choice of Qub.

\begin{theorem}
   Let $\alpha$ be a sentence whose quantum tree
    is the operator-sequence $(U_1,\ldots,U_{Height(\alpha)-1})$.
     Given a quantum
    computational model $Qub$, consider the
    quregister-sequence  $(\ket{\psi_1},\ldots,
    \ket{\psi_{Height(\alpha)}})$ that is determined
    by $Qub$ and by the syntactical tree of $\alpha$.
Then, $U_i (\ket{\psi_{i+1}})= \ket{\psi_i}$ (for any $i$ such that $1
\leq i < Height(\alpha)$).
\end{theorem}
\begin{proof}
Straightforward
\end{proof}

The quantum tree of $\alpha$ can be naturally
regarded as a {\it quantum circuit\/} that computes the
output $Qub(\alpha)$, given the input $Qub(p_1),\ldots, Qub(p_n)$
(where $p_1,\ldots,p_n$ are the atomic occurrences of $\alpha$).
In this framework, each $U_i$ is the unitary operator that describes
the computation performed by the $i$-th layer of the circuit.

\end{document}